# Relaxation and derelaxation of pure and hydrogenated amorphous silicon during thermal annealing experiments

F. Kail (1), J. Farjas (1), P. Roura (1), C. Secouard (2), O. Nos (3), J. Bertomeu (3), F. Alzina (4) and P. Roca i Cabarrocas (5)

(1) University of Girona

(2) CEA Grenoble

(3) University of  Barcelona

(4) Institut Català de Nanotecnologia (ICN)

(5) Ecole Polytechnique Palaiseau

The structural relaxation of pure amorphous silicon (a-Si) and hydrogenated amorphous silicon (a-Si:H) materials, that occurs during thermal annealing experiments, has been analysed by Raman spectroscopy and differential scanning calorimetry.  Unlike a-Si, the heat evolved from a-Si:H cannot be explained by relaxation of the Si-Si network strain, but it reveals a derelaxation of the bond angle strain. Since the state of relaxation after annealing is very similar for pure and hydrogenated materials, our results give strong experimental support to the predicted configurational gap between a-Si and crystalline silicon.

Hydrogenated amorphous silicon (a-Si:H) owes its high technological interest in applications such as photovoltaic energy conversion and flat panel displays to the beneficial role of hydrogen. Thanks to the Si-H groups of its structure, the high concentration of Si dangling bonds found in pure amorphous silicon (a-Si) is drastically reduced by either direct saturation by the H atoms or by relaxation of the Si-Si network strain induced by hydrogen [1]. Although experience tells us that the best electrical properties are achieved in those a-Si:H films where the relaxation degree is higher [2, 3], no studies have been devoted to compare the degree of strain relaxation induced by hydrogen incorporation in a-Si with that achieved by thermal annealing of pure a-Si.

When a-Si obtained by Si ion implantation in crystalline silicon (c-Si) is thermally annealed, it evolves spontaneously towards states of lower energy through a progressive diminution of the bond angle dispersion, $\Delta\theta$, around the tetrahedral angle of c-Si, and through point defect recombination [4, 5, 6]. It is said that the covalent network after annealing is "more relaxed".

In fact, according to the theoretical models of a-Si, there is a "relaxed state" of a-Si characterised by a minimum value $\Delta\theta_{relax} \approx 6.6º$ [7]. Below 6.6º, no models can be built, i.e. a configurational gap is predicted between the allowed microscopic configurations of a-Si ($\Delta\theta > 6.6º$) and c-Si ($\Delta\theta = 0$). This value has not been experimentally reached to date, and the question arises of whether this discrepancy is because the $\Delta\theta$ value of the relaxed state is higher than predicted, or it is due to kinetic reasons. Recently, we have shown that before crystallisation begins, structural relaxation processes stop near 8.7º because most of the mobile defects, whose diffusion facilitates network restructuring, have already recombined [8]. This conclusion leaves the door open to the possibility of reaching a higher degree of relaxation if alternative mechanisms for network restructuring were to come into play. In that case, further diminution of $\Delta\theta$ would be possible and, eventually, a value of $\Delta\theta$ below 6.6º could be reached if the real minimum value of $\Delta\theta$ were lower than predicted.

Although initial studies of a-Si relaxation focused on the evolution of $\Delta\theta$, it soon became clear that, in addition to dangling-bonds (DB), high densities (around $1 \cdot 10^{21}$ cm$^{-3}$ or $\approx 1$ defect per 100 Si-atoms) [4, 5] of non-paramagnetic electrically-active defects do exist in this material. Although most of them recombine during annealing [4, 9], our recent analyses [6, 8] indicate that the heat released during structural relaxation is mainly due to a reduction in bond-angle strain (i.e. a reduction in $\Delta\theta$).

In contrast to pure a-Si, whose structural relaxation has been extensively analysed, similar studies on hydrogenated amorphous Si (a-Si:H) are very scarce. It appears that the main structural transformations prior to crystallisation are driven by dehydrogenation and atomic hydrogen diffusion. Increments of DB [10, 11] and weak-bond (WB) [10] densities have been reported. On the other hand, rearrangement of the void structure [12] is probably the cause of the pronounced variations in hydrogen diffusivity observed during low-temperature annealing [13]. Finally, densification up to the mass density of pure a-Si is indicative of long-range rearrangements of the Si-Si network after dehydrogenation [14]. Probably, this extensive restructuring process, absent in pure a-Si, gives the materials further opportunities to reach the relaxed state and, consequently, allows us to address the fundamental problem of the minimum value of $\Delta\theta$ mentioned above.

The aim of this letter is to show that structural relaxation in a-Si:H is qualitatively different to that in a-Si both in terms of the structural transformations taking place and the microscopic mechanisms at work. We can draw these conclusions thanks to our Raman spectroscopy results, which serve to quantify the variations in bond angle dispersion before and after the heat treatment, and to the calorimetric experiments that monitor the heat exchanged during heating ramps.

A wide variety of a-Si and a-Si:H materials were obtained by various deposition techniques: a-Si films on glass by electron beam evaporation (EBE); a-Si:H films on glass by hot wire chemical vapour deposition (HWCVD); a-Si:H films (label "TF") on glass, a-Si:H flakes (label "F") collected from the inner walls of the reactor chamber and a-Si:H nanoparticles ("npSi") by plasma enhanced CVD (PECVD) (Table I). Several of these PECVD materials were grown under conditions where silicon clusters and nanocrystals are produced in the plasma and contribute to the growth (pm-Si:H). The hydrogenated films were deposited in conditions leading to device-quality material and all the samples were undoped. The hydrogen content of the samples, $n_H/n_{Si}$, was determined by elementary analysis (EA2400 Perkin Elmer). The annealing experiments were done by heating the samples in inert atmosphere at a constant rate (usually at 40 K/min) in a differential scanning calorimeter (DSC822 from Mettler Toledo) that delivers the heat exchanged as a continuous trace. Once the maximum temperature was reached, the sample was cooled down to room temperature without any isothermal stage. The dehydrogenation rate of selected samples during heating was measured by mass spectrometry (MKS Spectra Quadrupole of Micro Vision Plus). Finally, the material structure was analysed by Raman spectroscopy (micro-Raman DILOR) with an excitation wavelength of 632 nm and magnification of 50. Special care was taken to avoid local heating of the sample due to the laser beam. In particular, the anti-Stokes signal was recorded in selected samples to verify this condition.

Great attention has been paid to remove oxygen from the DSC furnace atmosphere because oxidation is enhanced during dehydrogenation of a-Si:H [15]. For a 'nominally' inert $N_2$ atmosphere, we always obtained a high exothermic signal extending up to the crystallization temperature ($\approx$ 700ºC) that was erroneously interpreted as a consequence of relaxation processes [16, 17]. This signal disappeared when more inert conditions were used (higher $N_2$ flow rates) revealing a less intense signal containing endothermic contributions (Fig. 1a).

In Fig. 1a we plot the DSC thermograms of two samples showing the heat evolved from the material before crystallisation begins, $Q_{relax}$. A small exothermic signal beginning at around 360ºC has been measured for the a-Si sample. Notice that, as expected, the signal threshold is close to the substrate deposition temperature (350ºC) used to deposit this particular sample. The area below the curve extrapolated up to 600ºC delivers a relaxation heat of $Q_{relax} = 40 \pm 20$ J/g. In contrast, the DSC signal of the hydrogenated sample has a more complex structure. Although it is very difficult to know with accuracy the absolute zero of the DSC signal, its comparison with the curve affected by residual oxidation makes clear that an endothermic contribution begins around 300ºC. This contribution has a dependence on temperature similar to that of the dehydrogenation rate (Fig. 1b).

To analyse the changes in the Si-Si covalent network caused by annealing, we measured the Raman spectra of the samples in their as-grown state and after the heat treatment.

Typical spectra are shown in Fig. 2. To highlight the slight variations in the TO-band width, the x-axis of each spectrum has been shifted so that the maxima are located at the origin. After annealing, the width of the pure a-Si samples remains unchanged or diminishes, whereas it increases for all the hydrogenated samples. The variations in the half width on the high wavenumber side ($\delta(\Gamma/2)$) are summarised in Table I. Finally, the values of $\Gamma/2$ were obtained by fitting the spectra with a set of Gaussian components [3]. The results are collected in Fig. 3 together with the evolution observed for a-Si films obtained by ion implantation [4]. The absolute values of $\Gamma/2$ depend on the particular fitting procedure (essentially the baseline choice but also the number of Gaussian components used). Since the same fitting procedure was applied to our spectra, we have confidence in the ordering of the values obtained for our samples, which has been verified by superposing the spectra as in Fig. 2. However, we consider that the values for ion-implanted a-Si in relation to our samples may contain an uncertainty of $\pm 1$ cm$^{-1}$. In addition, according to our analysis published in ref.[8] the temperature of these samples has been increased by 65ºC by to the published values [4] to take into account that anneals lasted 45 min.

In accordance with the widely accepted interpretation of the Raman TO band width, the values of $\Gamma/2$ have been converted to $\Delta\theta$ (right y-axis of Fig. 3) by using the particular relationship of Beeman et al. [7]:

$$\Gamma/2 (in\, cm^{-1}) = 7.5 + 3\Delta\theta (in \deg). \qquad (1)$$

It is clear from Fig. 3 that $\Delta\theta$ decreases with annealing for pure a-Si whereas it increases for the hydrogenated materials. In this context it is worth to comment that recent calculations [18] have analyzed at the microscopic level the effect of hydrogenation on the network strain. It is shown that hydrogen incorporation near small voids and divacancies tends to occur at the most strained bonds, thus reducing strain. It is thus expected that the strain ($\Delta\theta$) would increase after dehydrogenation.

Given that $\Delta\theta$ diminishes for a-Si, the interpretation of $Q_{relax}$ in these samples is straightforward. The bond-angle dispersion entails a built-in strain and, consequently, a strain energy that can be quantified as [19]:

$$U_{\Delta\theta} = A(\Delta\theta)^2, \qquad (2)$$

where A is a constant that depends on the Si-Si bond-force constants. We have recently shown [6] that the diminution of $U_{\Delta\theta}$ is the main contribution to the relaxation heat of ion-implanted a-Si and we obtained A $\approx$ 3.0 J/(g·deg$^2$). If we apply this value to the EBE350 sample, for which $|\delta\Delta\theta| \approx 0.8º$, a value around 25 J/g is predicted for $Q_{relax}$, which agrees with the area below the DSC curve in Fig. 1a ($40 \pm 20$ J/g).

In the case of hydrogenated samples, in addition to small residual oxidation, several processes may contribute to $Q_{relax}$: a) the increment of $\Delta\theta$, b) the increment of DB density and c) dehydrogenation. The first two processes are endothermic and weak. According to Table I and Eq. (2), the contribution of $\Delta\theta$ would be lower than 30 J/g for all the samples. On the other hand, we know that the DB density increases up to $5 \cdot 10^{19}$ cm$^{-3}$ upon annealing at high temperature [10], and this results in a small contribution of 4 J/g to the relaxation heat (formation energy of one DB $\approx$ 1eV [16]) that would explain the negative slope of the curve observed at high temperature (Fig. 1a). Finally, the small exothermic values reported for dehydrogenation ($\approx 50$ meV/H-atom [20]) would lead to contributions of about 20 J/g ($n_H/n_{Si} \approx 10\%$). This analysis explains why: a) the DSC curve is more complex than that of the pure a-Si samples, b) the DSC signal is very weak and c) weak exothermic contributions are expected.

The increment of $\Delta\theta$ in a-Si:H deduced from the Raman spectra is consistent with the increment of the WB density reported by other authors [10, 11]. In fact, WBs are highly strained Si-Si bonds and, consequently, they correspond to the tail of the $\theta$ distribution. When films are annealed in vacuum (or in inert atmosphere), most hydrogen is lost and the density of WBs (measured through the Urbach tail slope) increases [10]. However, when the experiment is done in hydrogen plasma, which prevents intense dehydrogenation, the WB density remains unchanged [11]. So, restructuring of the Si-Si network in a-Si:H is not directly related to the network temperature, but to dehydrogenation. The contrary holds for a-Si, whose structural relaxation is intimately related to the diffusion of network defects [8].

We have arrived, finally, at the interesting question posed in the introduction about the significance of the $\Delta\theta$ value obtained for the relaxed samples. First, it should be noted that, after annealing, we obtain $\Delta\theta$ values in the narrow range between 9.0 and 9.7º, irrespective of the deposition technique. Second, since few hydrogen atoms are left in the a-Si:H samples after annealing, it is tempting to interpret this value as the minimum value for any allowed configuration of pure a-Si. The fact that this state is achieved with the hydrogenated samples supports this assertion because after dehydrogenation a) 99.9% of the DBs left behind by the hydrogen atoms recombine and b) the material's mass density increases up to the value of pure a-Si [14]. This means that during annealing the Si-Si network undergoes extensive restructuring, thus allowing the material to explore all possible microscopic configurations. Consequently, if a-Si configurations of lower $\Delta\theta$ existed, they would be reached during network restructuring because of their lower energy. And finally, an independent experiment that demonstrates the high stability of the configuration with $\Delta\theta \approx 9-9.7º$ has been done with the a-Si EBE samples. Isothermal anneals lasting as long as 50 hours have not been able to reach lower values of $\Delta\theta$. Consequently, we think that our results provide strong experimental evidence for the existence of a configuration gap between c-Si and a-Si defined by $\Delta\theta \approx 9º$ [21]. Furthermore, since $\Delta\theta$ increases after dehydrogenation for the hydrogenated samples, our results show that, as expected, the presence of Si-H groups in the covalent network reduce the bond-angle strain and even reduce the configurational gap.

In summary, thermal annealing experiments on a-Si reveal a decrease of Si-Si network strain whereas an increase is observable in a-Si:H. In other words, the Si-Si network undergoes relaxation in a-Si but it 'derelax' in a-Si:H. Our experiments also provide strong support for the existence of the predicted configurational gap between a-Si(:H) and c-Si. This gap would decrease for hydrogenated material.

*Additional note:* during the revision of the manuscript, we have known that the WB density in a-Si:H films can be highly reduced by ion-bombardment during growth [22]. We wonder if $\Delta\theta$ in samples prepared in this way would be significantly lower than our reported values.

This work has been partially supported by the Spanish Programa Nacional de Materiales under contract number MAT2006-11144 and by the Generalitat de Catalunya under contract number 2005SGR-00666. One of us (FK) acknowledges financial support from the MEC (SB2006-0062).

**References**

[1] ZE. Smith and S.Wagner, Phys.Rev.Lett. **59**, 688 (1987).

[2] M. Brinza, G. Adriaesens and P. Roca i Cabarrocas, Thin Solid Films **427** 123 (2003).

[3] S. Vignoli, P. Melinon, B. Masenelli, P. Roca i Cabarrocas, A. M. Flank and C. Longeaud, J. Phys. :Condensed Matter **17** 1279 (2005).

[4] S. Roorda, W. C. Sinke, J. M. Poate, D. C. Jacobson, S. Dierker, B. S. Dennis, D. J. Eaglesham, F. Spaepen and P. Fuoss, Phys. Rev. B **44**, 3702 (1991).

[5] P. A. Stolk, F. W. Saris, A. J. M. Berntsen, W. F. van der Weg, L.T.Sealy, R. C. Barklie, G. Krötz and G. Müller, J. Appl. Phys. **75**, 7266 (1994).

[6] P. Roura, J. Farjas and P. Roca i Cabarrocas, J. Appl. Phys. **104,** 073521 (2008).

[7] D. Beeman, R. Tsu and M. F. Thorpe, Phys. Rev. B **32,** 874 (1985).

[8] P. Roura and J. Farjas, Acta Mater. **57**, 2098 (2009).

[9] J. H. Shin and H. A. Atwater, Phys. Rev. B **48**, 5964 (1993).

[10] K. Zellama, L. Chahed, P. Sladek, M. L. Theye, J. H. von Bardeleben and P. Roca i Cabarrocas, Phys. Rev. B **53**, 3804 (1996).

[11] N. H. Nickel and W. B. Jackson, Phys. Rev. B **51,** 4872 (1995).

[12] D. L. Young, P. Stradins, YQ. Xu, L. M. Gedvilas, E. Iwaniczko, YF. Yan, H. M. Branz, Q. Wang and D. L. Williamson, Appl. Phys. Lett. **90**, 081923 (2007).

[13] S. Mitra, R. Shinar and J. Shinar, Phys. Rev. B **42,** 6746 (1990).

[14] Z. Remes, M. Vanecek, A. H. Mahan and R. S. Crandall, Phys. Rev. B **56**, R12710 (1997).

[15] D. Das, J. Farjas, P. Roura, G. Viera and E. Bertran, Appl. Phys. Lett. **79**, 3705 (2001).

[16] P. Roura, J. Farjas, Ch. Rath, J. Serra-Miralles, E. Bertran and P. Roca i Cabarrocas, Phys. Rev. B **73**, 085203 (2006).

[17] P. Roura, J. Farjas and P. Roca i Cabarrocas, Thin Solid Films **517**, 6239 (2009).

[18] S. Chakraborty, D. A. Drabold, Phys. Rev. B **79,** 115214 (2009).

[19] T. Saito, T. Karasawa and I. Ohdomari, J. Non-cryst. Solids **50**, 271 (1982).

[20] J. Farjas, D. Das, J. Fort, P. Roura and E. Bertran, Phys. Rev. B **65**, 115403 (2002).

[21] Although this value is clearly higher than the theoretical 6.6º, the precise discrepancy depends on the precision of the $\Delta\theta(\Gamma/2)$ relationship used here (Eq. 1).

[22] G. Ganguly, I. Sakata and A. Matsuda, J. Non-Cryst. Solids. **198-200**, 300 (1996).

**Figure Captions.-**

Figure 1.- a) Heat evolved from of a pure a-Si sample (EBE350) and of a hydrogenated sample (aSiHW) measured in normal inert conditions and in a highly inert atmosphere to minimize residual oxidation (exo/endothermic contributions point up/downward) b) Dehydrogenation rate ($dn_H/dt$) of the hydrogenated sample..

Figure 2.- The width of the Raman TO band increases (decreases) during annealing for the a-Si:H (a-Si) samples.

Figure 3.- Evolution of the TO half-width during annealing and the corresponding dispersion of Si-Si bond-angles according to the Beeman formula (Eq. 1). For each sample, the lowest temperature is the deposition temperature, i.e. the point corresponds to the as-grown condition. The ion-amorphised a-Si points are taken from ref. [4].

**Table I.-** Raman TO half-width ($\Gamma/2$), its increment during annealing of the samples and the hydrogen content, $n_H/n_{Si}$

| sample | $\Gamma/2$ | $\delta(\Gamma/2)$ | $n_H/n_{Si}$ |
|---|---|---|---|
| | (cm$^{-1}$) | | |
| **a-Si:H grown by PECVD** | | | |
| aSi(F) | 34 | +0.5 | 0.12 |
| pmaSW(TF) | 34 | +1 | ≈0.15 |
| tfa(TF) | 34 | +0.8 | -- |
| pm(TF) | 33.4 | +2.2 | ≈0.15 |
| npSi | 34.5 | +1.5 | 0.28 |
| **a-Si:H grown by HWCVD** | | | |
| pSiHW | 34.7 | +1.0 | -- |
| aSiHW | 34.0 | +2.5 | 0.13 |
| **a-Si grown by EBE** | | | |
| EBE350 | 38.3 | -2.5 | 0 |
| EBE425 | 37.5 | -1.5 | 0 |
| GRE350 | 35.5 | 0 | 0 |
| GRE425 | 36.8 | -1.7 | 0 |
| GRE500 | 35.3 | 0 | 0 |

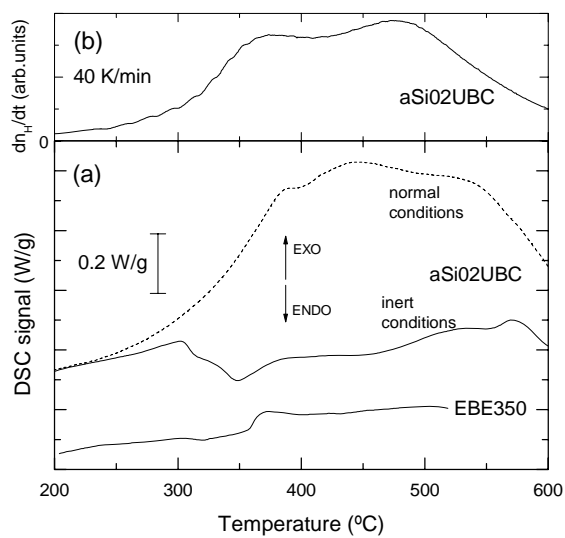

Figure 1.- a) Heat evolved from of a pure a-Si sample (EBE350) and of a hydrogenated sample (aSiHW) measured in normal inert conditions and in a highly inert atmosphere to minimize residual oxidation (exo/endothermic contributions point up/downward) b) Dehydrogenation rate ($dn_H/dt$) of the hydrogenated sample..

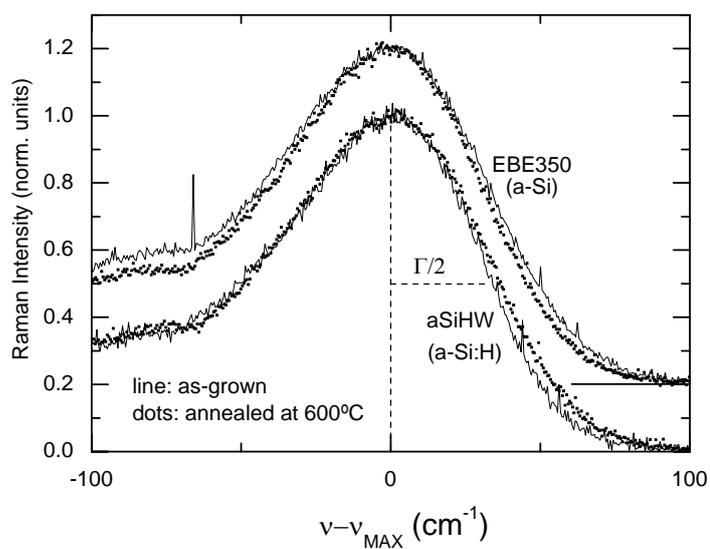

Figure 2.- The width of the Raman TO band increases (decreases) during annealing for the a-Si:H (a-Si) samples.

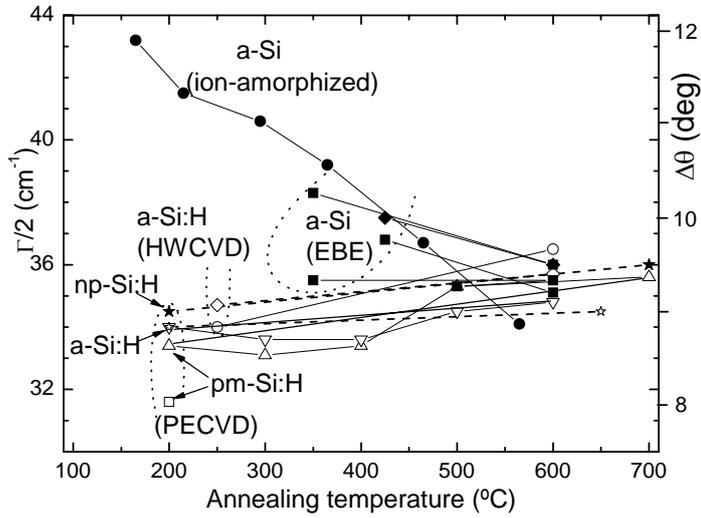

Figure 3.- Evolution of the TO half-width during annealing and the corresponding dispersion of Si-Si bond-angles according to the Beeman formula (Eq. 1). For each sample, the lowest temperature is the deposition temperature, i.e. the point corresponds to the as-grown condition. The ion-amorphised a-Si points are taken from ref. [4].